\documentclass[notitlepage,nofootinbib,showpacs,preprintnumbers,superscriptaddress,amsmath,amssymb,12pt]{revtex4}
\usepackage{graphicx}
\usepackage{dcolumn}
\usepackage{bm}
\usepackage{amsmath}
\usepackage{multirow}

\newcommand{\mean}[1]{\mbox{$\langle{#1}\rangle$}}

\usepackage[colorlinks=true]{hyperref}

\begin{document}

\title{ Microlens Array Laser Transverse Shaping Technique for Photoemission Electron Source} 
\author{A. Halavanau$^{1,5}$, G. Ha$^{2,3}$, G. Qiang$^{2,4}$, W. Gai$^2$, J. Power$^2$, P. Piot$^{1,5}$, E. Wisniewski$^{2}$, \\
D. Edstrom$^5$, J. Ruan$^5$, J. Santucci$^5$, \\
$^1$ Department of Physics and Northern Illinois Center for Accelerator \& \\
Detector Development, Northern Illinois University, DeKalb, IL 60115, USA \\
$^2$ Argonne Wakefield Accelerator, Argonne National Laboratory, Lemont, IL, 60439, USA\\
$^3$ POSTECH, Pohang, Kyoungbuk, 37673, Korea\\
$^4$ Accelerator laboratory, Department of Engineering Physics, Tsinghua University, Beijing, China\\
$^5$ Fermi National Accelerator Laboratory, Batavia, IL 60510, USA} 
\date{\today}
\begin{abstract}
A common issue encountered in photoemission electron sources used in electron accelerators is distortion of the laser spot due to non ideal conditions at all stages of the amplification. 
Such a laser spot at the cathode may produce asymmetric charged beams that will 
result in degradation of the beam quality due to space charge at early stages of 
acceleration and fail to optimally utilize the cathode surface. In this 
note we study the possibility of using microlens arrays to dramatically improve 
the transverse uniformity of the drive laser pulse on UV photocathodes at both Fermilab Accelerator Science \& Technology (FAST) facility
and Argonne Wakefield Accelerator (AWA). 
In particular, we discuss the experimental characterization of 
the homogeneity and periodic patterned formation at the photocathode. 
Finally, we compare the experimental results with the paraxial analysis, ray tracing and wavefront propagation software.

\end{abstract}
\preprint{FERMILAB-TM-2634-APC}
\pacs{ 29.27.-a, 41.75.Fr, 41.85.-p, 42.15.Dp, 42.15.Eq, 42.30.Lr, 42.60.Jf}
\maketitle

\section{Introduction}
Photoemission electron sources are widespread and serve as backbones to a number of applications including high-energy particle accelerators, accelerator-based light sources, and ultra-fast electron diffraction setups. For a given photoemission electron-source design, the electron-beam properties, and notably its brightness, are ultimately limited by the initial conditions set by the laser pulse impinging the photocathode. A challenge common to most applications is the ability to produce an electron beam with uniform transverse density. Non uniformities in the transverse electron-beam density can lead to transverse emittance dilution or intricate correlations. 
Producing and transporting a laser pulse while preserving an homogeneous transverse density require a great amount of work. For instance the ultraviolet (UV) laser pulses usually required for photoemission from metallic or semiconductor cathode require the use of nonlinear conversion process to form the UV pulse from an amplified infrared (IR) pulse. 
This conversion process often introduces ``hot'' spots. 

In this paper we investigate an alternative technique and demonstrate the use of microlens arrays (MLAs) to directly homogenize the UV laser pulse. MLAs are fly's eye type light condenser often employed as optical homogenizer for various applications~\cite{mlaabcd,dickey2000laser,deOliveira}. 

Qualitatively, the principle of the MLA lies in redistributing the incoming light intensity across the light  beam spot. Typically MLAs are arranged in pairs. After passing through the MLAs, the light rays are collected by a ``Fourier" lens which focuses parallel rays from different beamlets to a single point at the image plane. Under proper conditions (distance to the Fourier lens and its focal length), the process leads to transverse homogenizing of the beam; see Fig.~\ref{drawing}. Therefore the MLA homogenization scheme is quite simple and appealing in the context of photocathode drive lasers.  

Alternatively, imaging the object plane of the single microlenses in the MLA with a ``Fourier" lens produces 
a set  of optical beamlets arranged as arrays (with pattern mimicking the microlens spatial distributions). 
The latter  configuration, is also relevant to electron sources as it can lead to the formation
 of transversely-segmented beams  with applications to beam-based diagnostics of 
accelerator (alignment, nonlinearities), single-shot quantum-efficiency map measurement, and coherent light sources in the THz regime and beyond. 

After briefly summarizing the principle of the MLA setup, we demonstrate its possible use to homogenize the ultraviolet (UV) laser spot of the photocathode drive laser at Fermilab Accelerator Science \& Technology (FAST) facility \cite{asta}. We establish the usefulness of MLAs to control the electron beam distribution in a series of experiments carried out at the Argonne Wakefield Accelerator (AWA) facility~\cite{manoel}.

\section{ABCD formalism}

\subsection{MLA ray-tracing considerations}
\begin{figure}[b]
\begin{center}
 \includegraphics[width=0.87\linewidth]{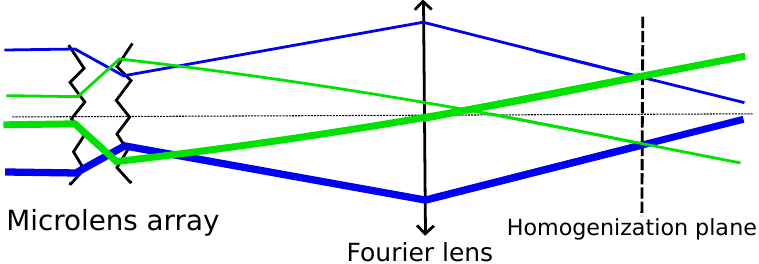}
 \caption{\label{drawing} Schematics of the microlens array configuration. Initial incoherence in the beam (thin/thick ray) becomes
evenly distributed at the homogenization plane.}
\end{center}
\end{figure}
We now analyze the typical MLA setup diagrammed in Fig.~\ref{drawing} to derive a few salient features relevant to homogenization using the {\sc abcd} formalism~\cite{mlaabcd}. We consider an initial ray to be characterized by the vector $(x_0, \alpha_0)$, where $x_0$ and $\alpha_0\equiv\frac{dx_0}{dz}$ are respectively the initial ray position and divergence (here, $z$ represents the path-length along the optical transport). As a simple example we consider a rectangular array of microlens in the $(x,y)$ plane with equal pitch in both transverse directions. Using the  {\sc abcd} formalism, and considering that the ray is within the aperture $\rho$ of  the lens with center located at $(x=mp,y=np)$, we can describe the  MLA with the linear transformation 
\begin{eqnarray}
\left( \begin{array}{c}
x_{1}-mp  \\
\alpha_{1}  \end{array} \right) =  \left( \begin{array}{cc}
1 & 0  \\
-1/f_2 & 1  \end{array} \right) \left( \begin{array}{cc}
1 & s  \\
0 & 1  \end{array} \right) \left( \begin{array}{cc}
1 & 0  \\
-1/f_1 & 1  \end{array} \right) \left( \begin{array}{c}
x_0-mp  \\
\alpha_0  \end{array} \right), 
\end{eqnarray}
where $(x_{1},\alpha_{1})$ is the beam vector after two MLA plates, $s$ is the spacing between two plates, $p$ is the array pitch, $f_1$
and $f_2$ are the focal lengths of the first and second microlens respectively. It 
should be pointed out that the ray initial and final 
coordinate satisfy $\sqrt{(x_0-mp)^2+(y_0-np)^2}\le \rho$ with $u \in 0,1$. Finally $n$ and $m$ indices are 
integer numbers that describe the position of each micro-lens in term of the pitch. Then, the 
output ray from the MLA setup can be further propagated up to the homogenization plane as 
\begin{eqnarray}
\left( \begin{array}{c}
x_{h}  \\
\alpha_{h}  \end{array} \right) =  \left( \begin{array}{cc}
1 & L  \\
0 & 1  \end{array} \right) \left( \begin{array}{cc}
1 & 0  \\
-1/F & 1  \end{array} \right) \left( \begin{array}{cc}
1 & d  \\
0 & 1  \end{array} \right) \left( \begin{array}{c}
x_1  \\
\alpha_1  \end{array} \right), 
\end{eqnarray}
where $(x_{h},\alpha_{h})$ is the ray vector at the homogenization plane, $d$ the distance between the Fourier lens and the MLA, $F$ the focal length of the Fourier lens and $L$ is the distance to the homogenization plane. 

From the formalism above one can deduce a few useful expressions. We consider the case when the two MLAs are identical ($f_1=f_2=f$) and located in the object plane of the Fourier lens ($L=F$). We further assume that there is no cross-talk between the
 microlens and their transformation only affects rays within a finite aperture smaller than the array pitch $\sqrt{(x_0-mp)^2+(y_0-np)^2}\le p/2$. Under these assumptions, we find the diameter of the image at the homogenization plane to be  
\begin{eqnarray}
\label{hom_spot}
 D_h \approx \frac{F p}{f^2}(2f-s)
\end{eqnarray}
in the limit of small ray divergence (as indicated by the independence of the equation on $d$).
For practical purposes, we also calculate the diameter of the beam at the Fourier lens plane to be 
\begin{eqnarray}
\label{lens_apt}
 A_F \approx \frac{d p}{f^2}(2f-s).
\end{eqnarray}
The latter equation is useful to estimate the required aperture. 

In practical simulations, the assumption $L=F$ might be challenging to satisfy. In such cases, the following expression is useful
to determine the beam size at a given location:
\begin{eqnarray}
D \approx \frac{p L}{f^2}(2f-s)+\frac{d p (2f-s)}{f^2}\frac{F-L}{F}. 
\end{eqnarray}

If the location $L$ is close to the focal plane, the resulting image will stay highly homogenized due to the finite size of the Airy disk. Moving away from the focal plane increases the density modulations. 

\subsection{Optical transport design}

Photoinjector setups generally incorporate relatively long (multi-meter scales) optical transport lines. The optical lines include transport from the laser room to the photoinjector enclosure (generally performed in the air or in moderate vacuum pipe) and the injection in the ultra-high-vacuum accelerator beamline up to the photocathode. Consequently, it is necessary to devise an optical transport line capable of imaging the homogenized laser profile on the photocathode surface. A commonly-used imaging setup, 
known as $4f$-imaging, is difficult to implement in the present case as it would require some of the lenses to be located in the vacuum chamber, as the ``imaging'' plane has to be much farther downstream than the ``object'' plane upstream.

However imaging can be achieved in numerous ways while accommodating the various constraints related to MLAs (limited apertures, 
available focal lengths, etc...). To construct the appropriate optical line, to start, we impose the vector of a ray in the homogenization plane $(x_h,\alpha_h)$ to be transported to a downstream imaging plane $(x_I,\alpha_I)$ via 
\[\left( \begin{array}{c}
x_I  \\
\alpha_I  \end{array} \right) = \mathbf{M} \left( \begin{array}{c}
x_h   \\
\alpha_h \end{array} \right), \quad \mbox{with~~}\mathbf{M}=\left( \begin{array}{cc}
{\cal M} & 0  \\
0 & 1/{\cal M}\end{array} \right),\]
where the magnification ${\cal M}$ is set to 1 for one-to-one imaging. The latter linear system yields four equations; an additional
constraint comes from the total length of the imaging transport. Therefore, the problem has 5 unknowns in total with some flexibility within available lenses.
Hence, it is possible to construct five-lens solution with distances between lenses as free parameters to make 
the corresponding system of linear equations well-defined.

\begin{figure}
\begin{center}
 \includegraphics[width=0.97\linewidth]{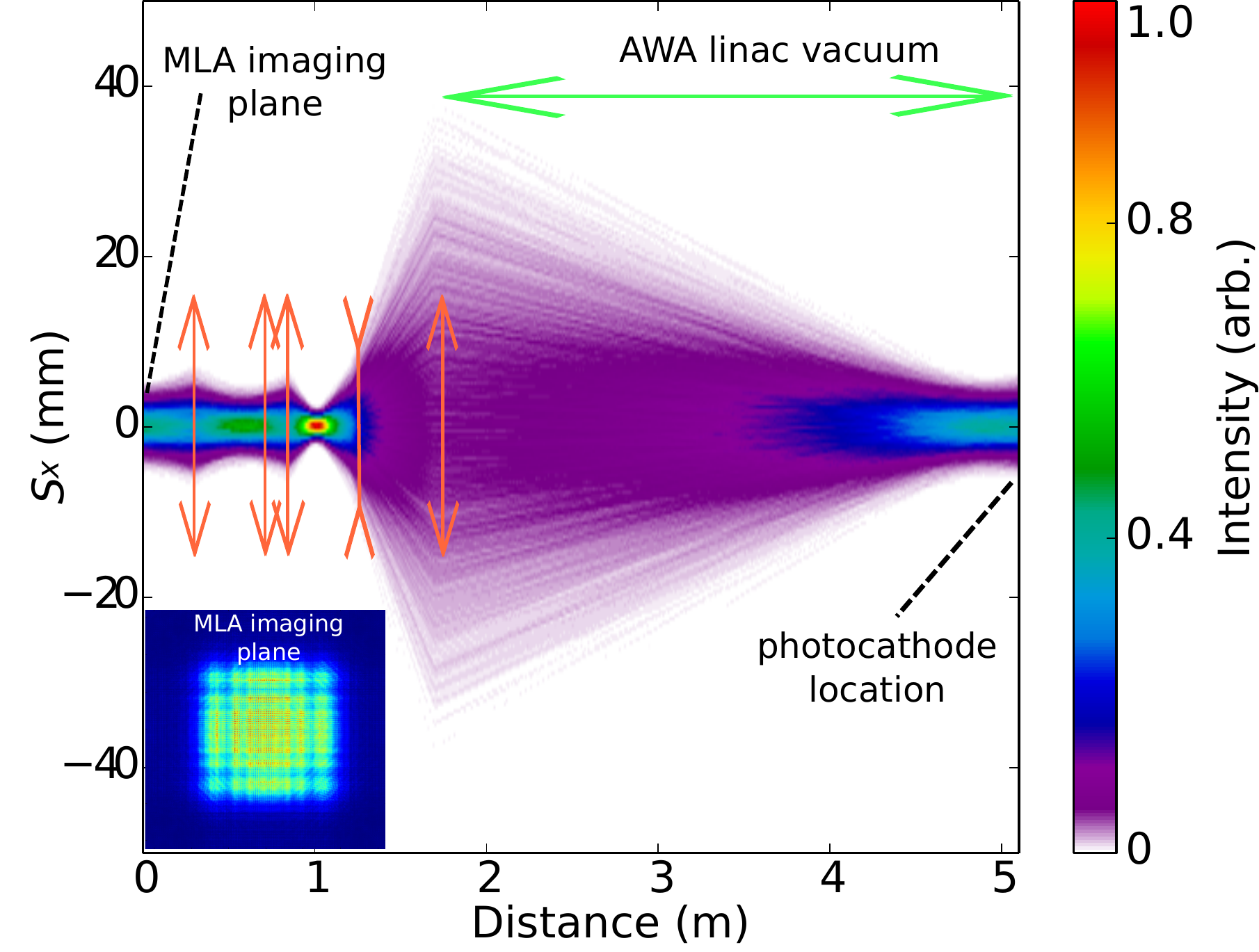}
 \caption{\label{image_transport} Gaussian ray-tracing of a five-lens transport capable of imaging the homogenized beam to the photocathode (this system was built at AWA).  The inset shows of a $5 \times 5$ simulated through the MLA using {\sc srw} (bottom left).}
\end{center}
\end{figure}

Such a five-lens system was designed and the beam evolution is depicted via Gaussian distributed ray-tracing in Fig.~\ref{image_transport}.
As it can be seen, the last lens aperture is a limiting factor, as the beam size in the setup under consideration is 
large before it slowly converges to the photocathode, yet it stays under the aperture limit of commercially available lenses. Note that
 the mirror inside the vacuum tube can result in beam clipping, therefore this transport 
line could be further optimized. Transverse instabilities coming from shot-to-shot 
change in the inhomogeneous beam distribution shown in Fig.~\ref{noarray}, for example, may result in charge fluctuations if the laser beam is collimated by an iris upstream of the MLA. To alleviate this possible issue we introduced a beam reducer in front of the MLA. 
\section{Optical measurements}
To evaluate the performance of the proposed scheme, we use two MLA's on the photocathode drive laser of the AWA \cite{manoel}. The input UV ($\lambda = 248$~nm) 
laser pulse is obtained from frequency tripling of an amplified IR pulse originating from Ti:Sp laser system.
 Downstream of the frequency tripler the UV pulse is further amplified by a two-pass excimer amplifier before transport to the accelerator vault. The laser is collimated and sent to the MLA setup described in Fig.~\ref{drawing} followed by the optical transport line shown in Fig.~\ref{image_transport}.  

\begin{figure}
 \includegraphics[width=0.97\linewidth]{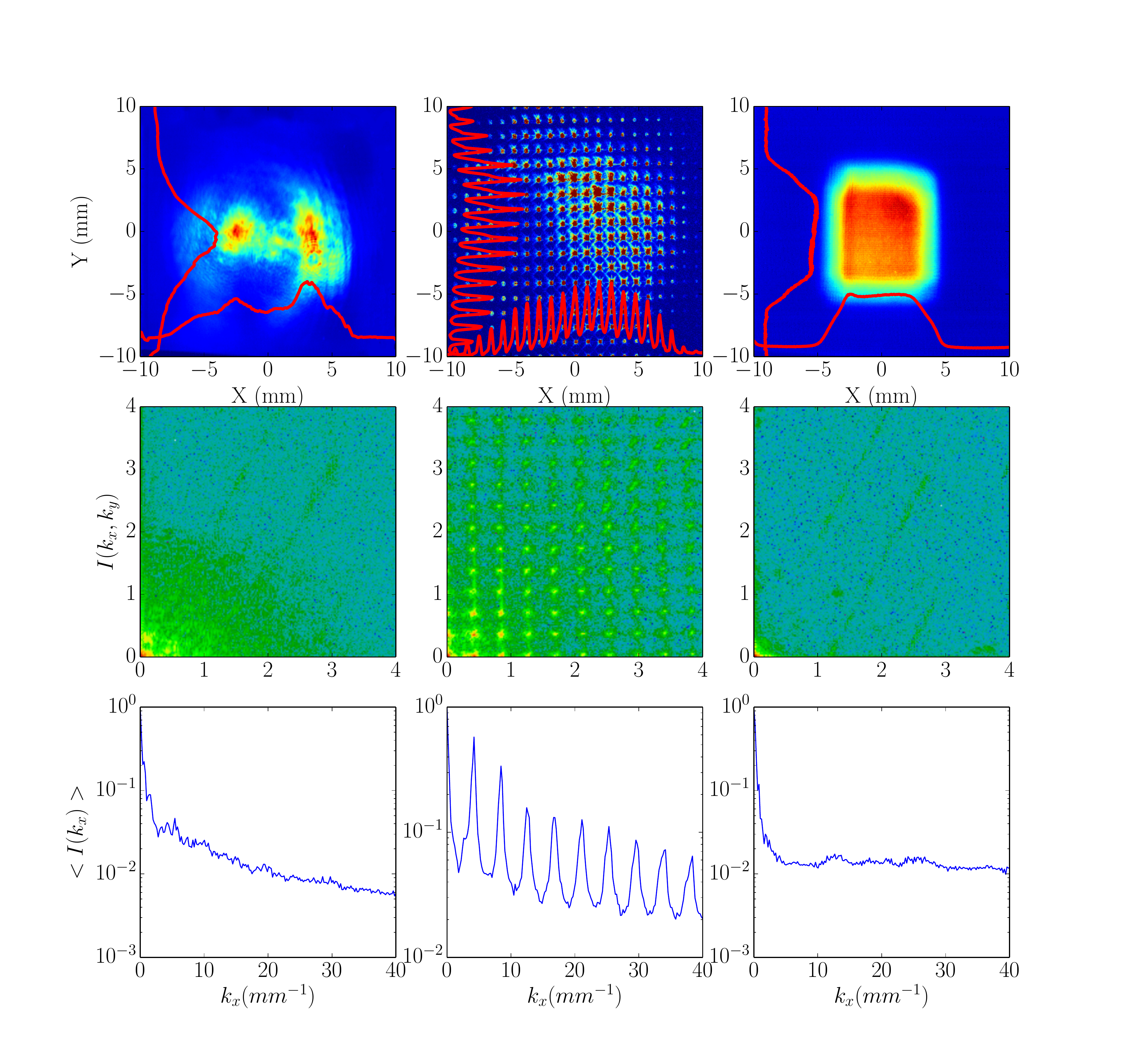}
 \caption{Measured UV laser without MLA (left column) and with MLA setup to produce beamlets (middle column) or as a homogenizer (right column). The upper, middle, and lower rows respectively 
correspond to the laser transverse density distribution, its 2D FFT, and the projected  spectrum along the horizontal frequency $k_x$.\label{noarray}}
\end{figure}

To gain confidence in the performances of the MLA setup, we performed several preliminary experiments. 
The first measurement was checking the consistency of the image with non perfectly collimated incoming laser beam. 
As it can be deduced from Fig.~\ref{drawing} the homogenization effect can be achieved even if the incoming beam has small a divergence. There is a critical value of beam divergence that causes crosstalk 
after the MLA and results in light loss $\tan \theta=p/2f$ \cite{sussinfo}. If the beam 
has initial uneven divergence along the $x$ and $y$ directions, we would observe a distortion 
of the square-shaped homogenized distribution in the homogenizing plane.  of the image.
In our experiment we observed only minor distortions; see Fig.~\ref{noarray}.
 This effect 
can be addressed more carefully in Fourier analysis of the laser profile.

Overall, we have observed a good agreement with Eq. \ref{hom_spot} and Eq. \ref{lens_apt}. Note, that the Fourier lens in the experimental setup should be placed at the distance $D>F$ from the array, where $F$ is the focal length of the Fourier lens. 

The second set of experiments consisted of quantifying the degree of beam homogenizing. 
This topic was explored in great detail in \cite{FZhou,Rihaoui}, here we will follow another route. 
The nominal UV laser pulse was used as a starting condition; see Fig. \ref{noarray} (left).  The transverse modulation can be quantified by considering the spatial Fourier transforms. Correspondingly, we consider the 
image $I(i,j)$ array  (where $i$ and $j$ are indexes associated to the horizontal $x$ and vertical $y$ directions) and consider its two-dimensional Fourier transform $\tilde{I}(k_i,k_j)$ performed using {\sc python} {\sc numPy} toolbox.
 Let's introduce the quantity $\mean{I(k_i)}$ which is the projection of 2D-intensity $I(i,j)$.
Fig.~\ref{noarray} (right image/plot) shows the results of the 2D Fourier transform
 (top) and the projection along the horizontal $k_i$ axis of the laser spot that displays
 typical microstructures observed in previous runs at FAST and AWA.  The spectrum displays some 
small modulations at low frequencies ($k_i<5$~mm$^{-1}$) that quickly decrease with frequencies to values $\le 5\times 10^{-2}$. 

\begin{figure}
 \includegraphics[width=0.95\linewidth]{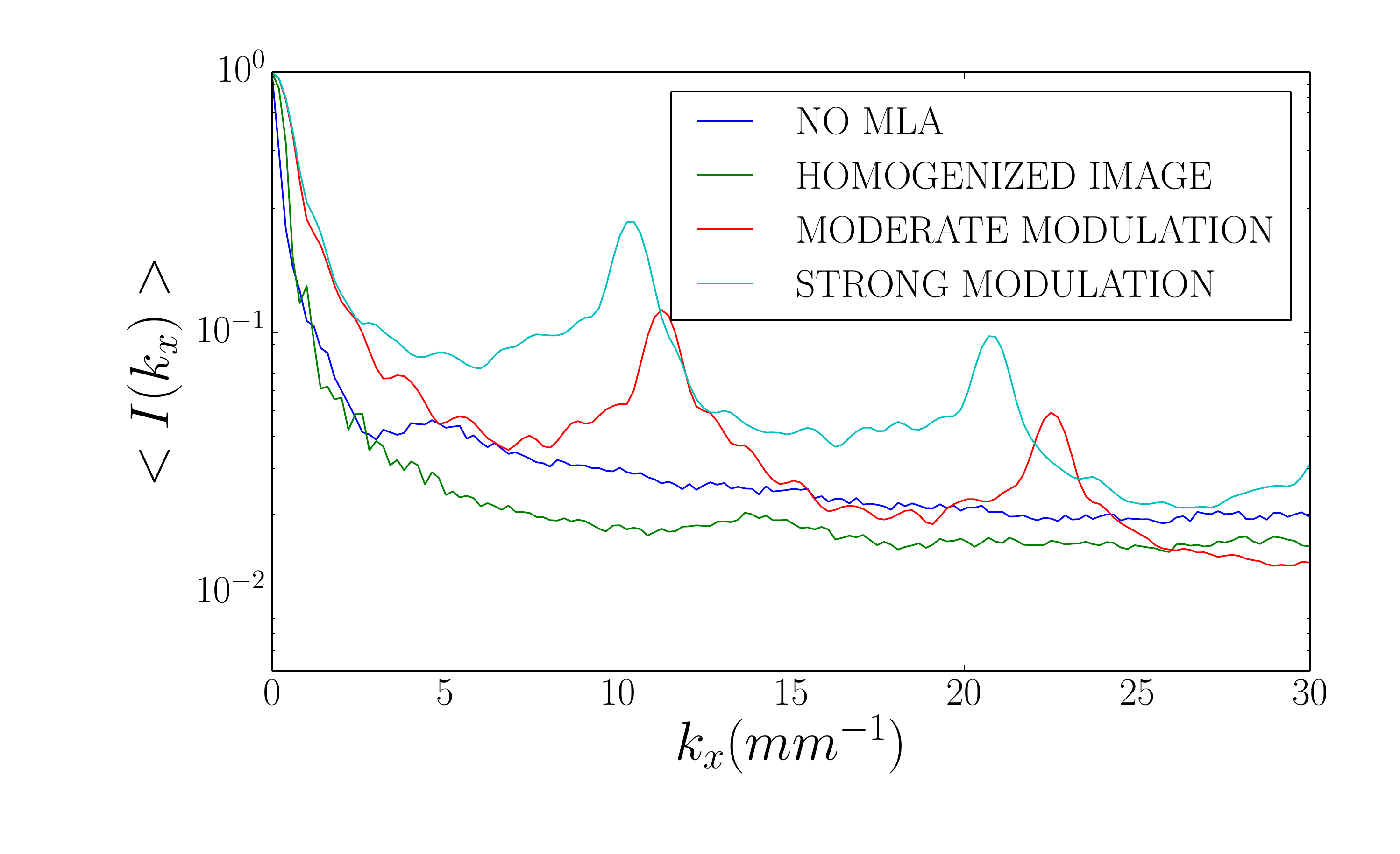}
 \caption{Averaged FFT spectrum for the different positions of the Fourier lens.\label{MLA_FFT}}
\end{figure}

When the MLA is set to homogenize the beam (see Fig. \ref{MLA_FFT}) the Fourier transform indicates that the low frequency modulations 
seen in the original beam are suppressed but high-frequency modulation is present at $k_i>12$~mm$^{-1}$. This modulation has 
a bunching factor on the order of $10^{-2}$.
Note that overall, the modulation wavelength decreases and at high degree of homogenization it will be fully dominated by
impurities on the photocathode emission surface, see Fig.~\ref{MLA_FFT}.

In the case when the MLA is set up to form a transversely modulated beam, the spectrum display high bunching factor at frequency and larger than the characteristic frequency associated to the total beam size; see Fig. \ref{MLA_FFT}. 

Overall, the described FFT spectrum behavior represented in combined on Fig. \ref{MLA_FFT}. Each curve is the averaging of 5 images taken after $f=250$ mm Fourier lens at  250 mm and further to study the off-focal modulation and pattern formation. The plot shows that in homogenization regime MLA setup significantly improves the image spectrum by suppressing the original modulations in the beam.

Finally, we quantified the laser power loss in the employed setup. The MLA plates did not have any UV anti-reflection (AR) coating, hence the power loss was $\sim 5$\% per surface or $\sim 20$\% per two MLAs. 
Additionally, the absence of AR coating on the UV lenses surfaces further introduce a power loss of $\sim 10$\% per lens.
Yet in many facilities the power budget of the laser is large so it is affordable to operate at a reasonable 
charge with the MLA and imaging transport system. In our experiment the laser power after passing through the
MLA setup and optical transport was measured to be 1.3~mJ, while the laser power upstream of the MLA was
measured to be 4.2~mJ.

\section{Electron beam simulations}

\begin{figure}
\label{beamline}
\centering
\includegraphics[width=1\linewidth]{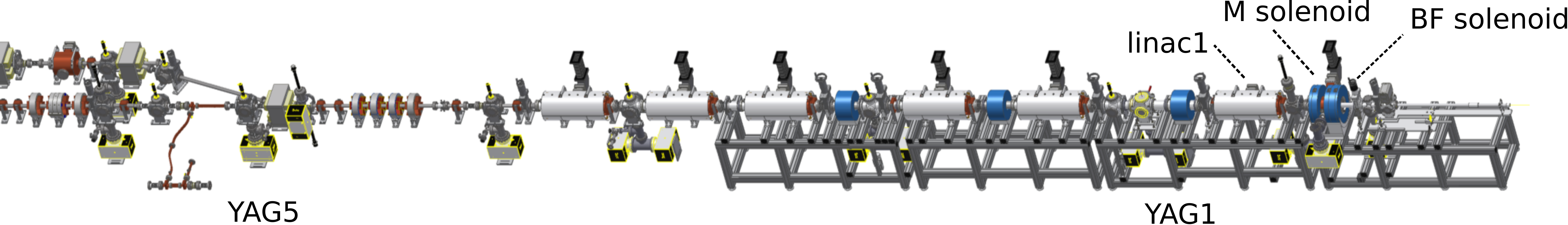}
\caption{AWA beamline overview (the beam direction from right to left). Bucking-focusing 
(BF) and matching (M) solenoids were adjusted to image the beam on YAG screens. The energy gain of one accelerating cavity (linac) is 10~MeV.}
\end{figure}

The electron-beam simulations and measurements were performed at AWA facility linac diagrammed in Fig.~\ref{beamline}; for a detailed description of the facility see Ref.~\cite{manoel}. The manipulated laser pulse impinges a high-quantum efficiency cathode located in a L-band RF gun to produce an electron bunch. 

The laser profile was used in numerical simulations to assess whether the modulation is within the bandwidth of possible amplification via collective effects (e.g. implying transverse space
 charge modulations that will eventually convert into energy modulations) or if they are simply smeared out via thermal-emittance effect as the beam is photo-emitted.

\begin{figure}
\centering
 \includegraphics[width=0.5\linewidth]{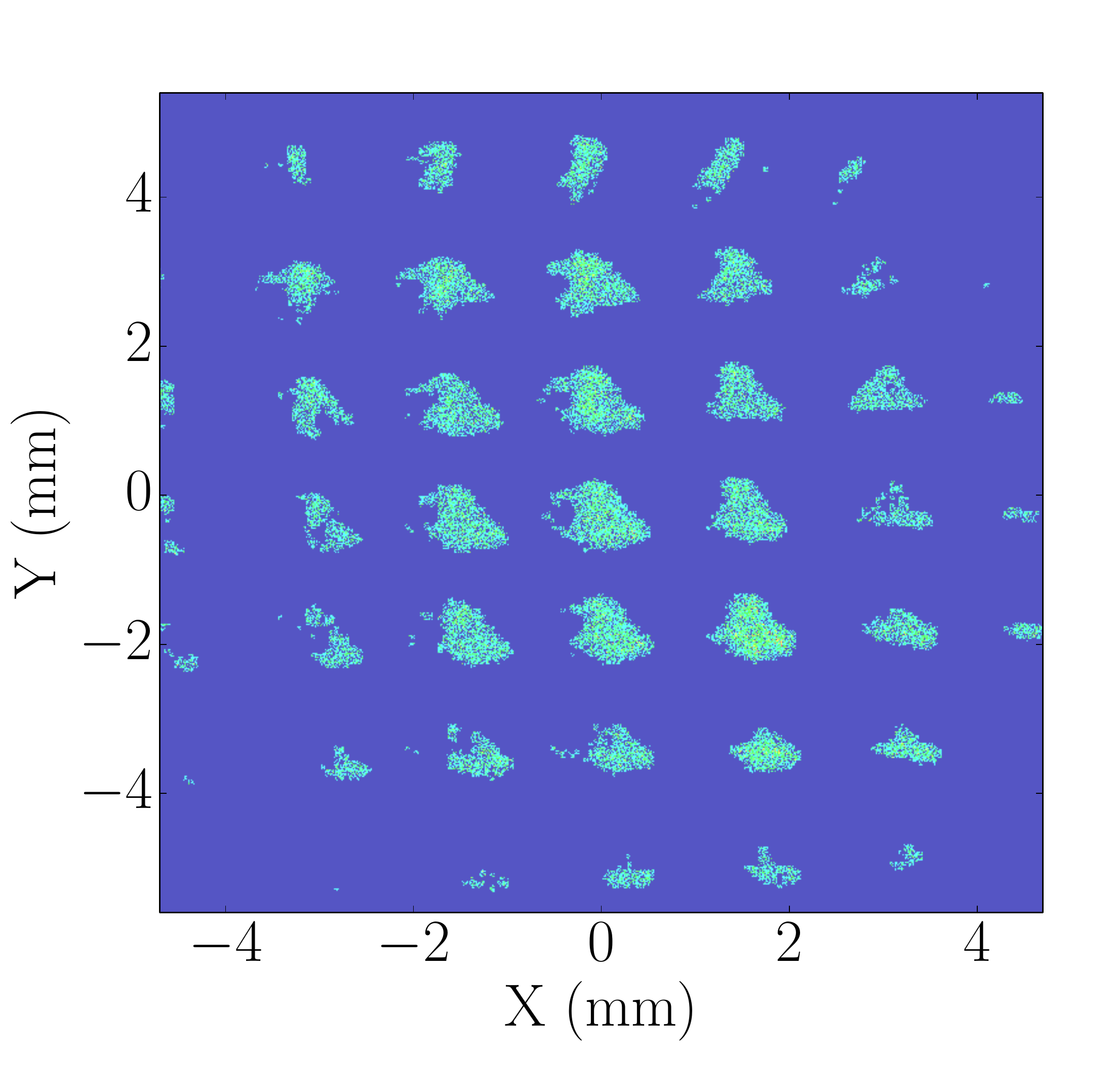}
   \caption{\label{GPTspot}Modulated laser spot converted into GPT particle distribution. }
\end{figure}

A code that converts laser spot picture into a particle distribution, was implemented~\cite{Rihaoui}. In our approach the measured laser-distribution image is taken as a tabulated two-dimensional distribution function that is used by a Monte-Carlo generator to produce the corresponding macroparticle distribution. The temporal laser distribution is 
taken to be Gaussian with rms $\sigma_t=2.5$~ps, consistent with streak camera measurements of the laser duration. The initial longitudinal component and transverse momentum are generated using  the {\sc astra} distribution
 generator~\cite{Astramanual} and consider a cathode with excess kinetic energy of 0.5~eV. The resulting particle distribution is
 then be easily converted to a required format for use in several beam-dynamics programs (Fig. \ref{GPTspot}).

\begin{figure}
\centering
 \includegraphics[width=0.57\linewidth]{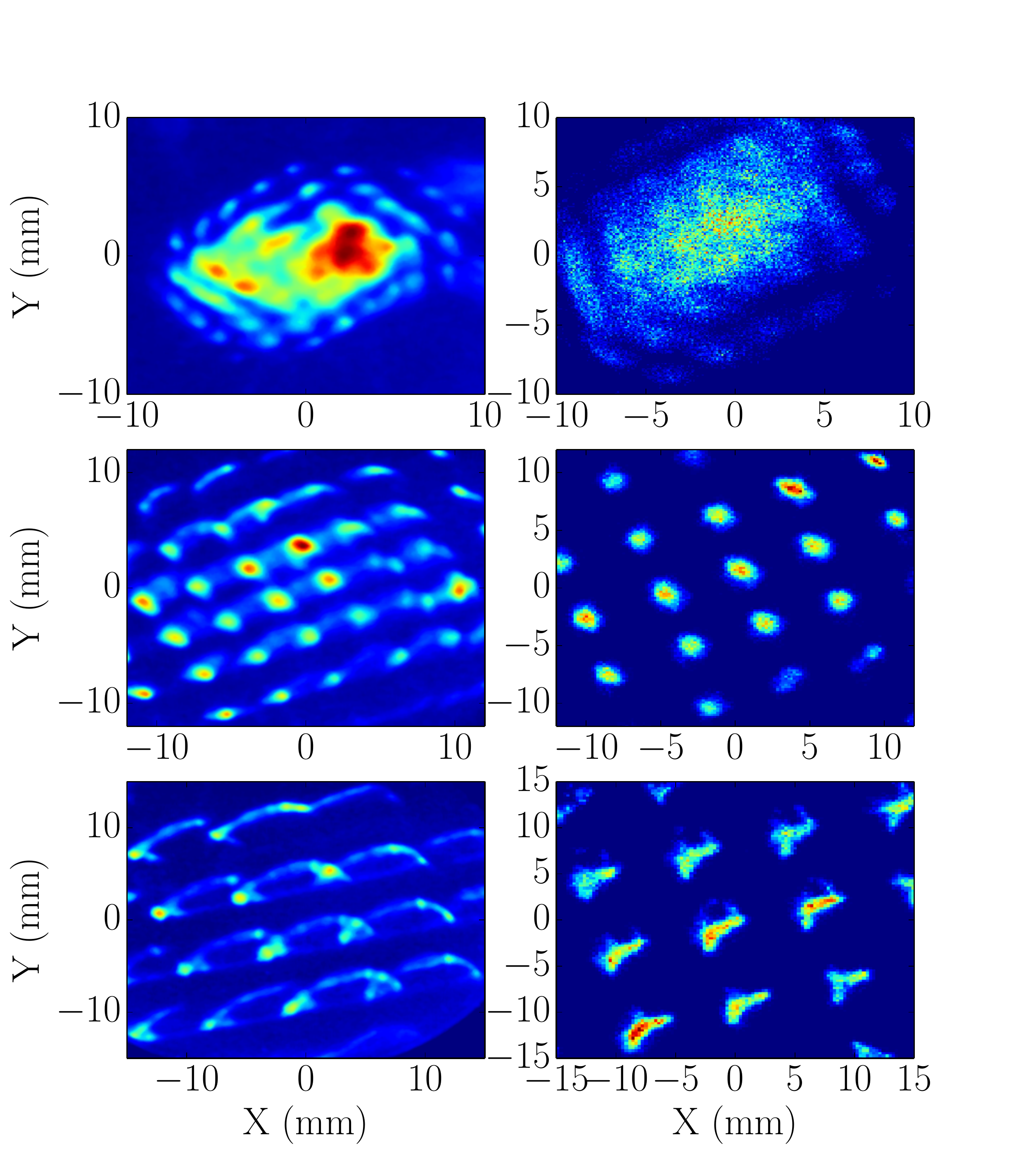}
 \caption{\label{simpat}7 MeV electron beam patterns (left) and GPT simulations (right) for the case of weak (215 A), medium (230 A) and strong (290 A) matching
solenoid current. }
\end{figure}

We carried out several simulations using the particle tracking code {\sc gpt}~\cite{GPT} to especially 
explore the impact of the MLA-homogenized beam on the resulting emittance and  possible formation and transport of patterned beams. 
One investigation also concerned the impact of space charge on the preservation of the modulation. The space charge 
force at early stages of acceleration was calculated via Barnes-Hut (BH) algorithm \cite{Barnes:1986} 
implemented in {\sc gpt}. This algorithm was tested to 
reproduce correct transverse and longitudinal beam dynamics in strong space-charge cases \cite{Halavanau:NIMA}.
{\sc gpt} model also shows good agreement with experimental results for low energy patterned beam experiment; see Fig.~\ref{simpat}.

\begin{figure}
\centering
\includegraphics[width=0.65\linewidth]{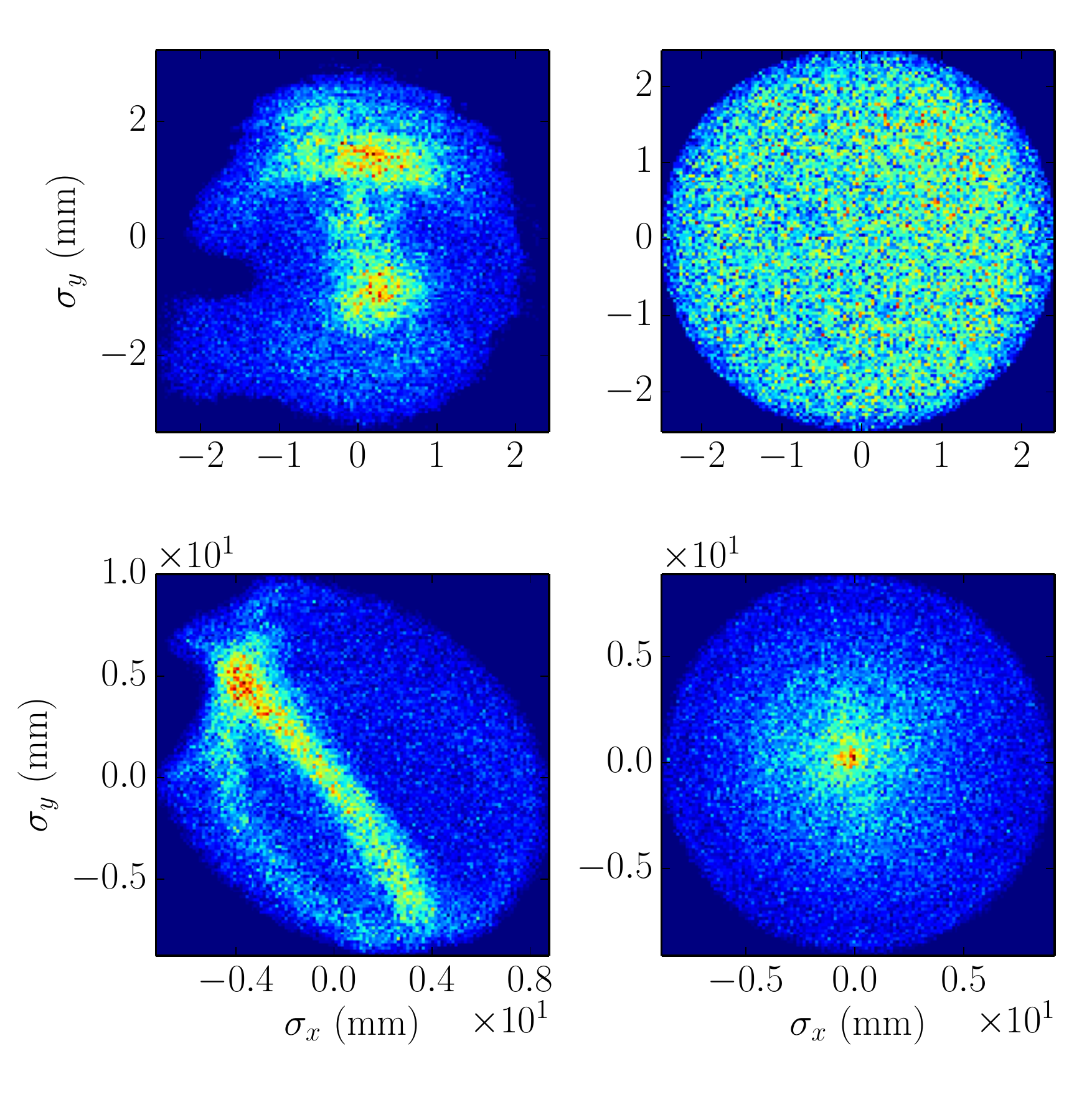}
\includegraphics[width=0.65\linewidth]{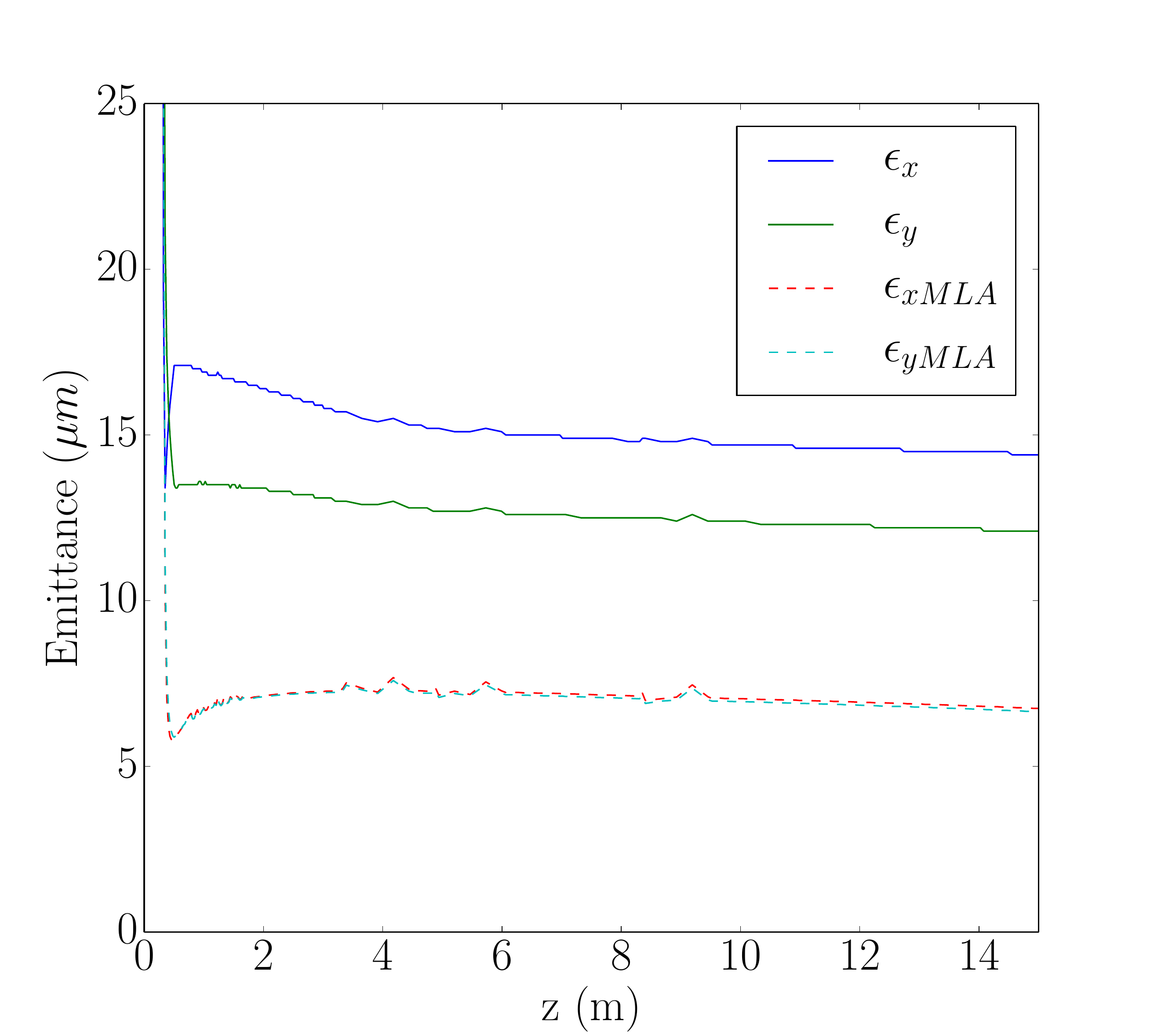}
  \caption{\label{beamemit}GPT beam emittance simulations. First row: electron distribution at the photocathode for no MLA (left) and MLA case (right). Second row:
  electron distribution at YAG5 for no MLA (left) and MLA case (right). Bottom plot: corresponding beam emittance at distance $z$ from the photocathode. }
\end{figure}

In order to investigate the impact of the MLA setup to homogenize the beam we 
consider the nominal and homogenized laser distribution (Fig.~\ref{beamemit}, upper row). 
To ensure a fair comparison, the total charge for both cases of distributions was set to 1~nC. Likewise, 
the RMS size of the distribution was fixed to 3~mm in both transverse direction (see Fig. \ref{beamemit}). 
The beam was transported through the nominal AWA lattice. 
Fig.~\ref{beamemit} (middle row) compares the beam distribution at 13~m from the cathode surface where the beam has reached an energy of 50~MeV. 
The beam asymmetric charge localization seen for the nominal case (Fig.~\ref{beamemit}, left column) disappears
 and the distribution becomes cylindrically symmetric when the MLA setup homogenizes the laser distribution (Fig.~\ref{beamemit}, right column) . 

Transverse modulations at the cathode eventually dissipate and lead to phase-space dilution or can affect 
the beam dynamics at higher energies. Thus, employing homogenized laser spot distributions can increase the beam quality. 
This is supported by our simulations which demonstrate that the beam emittance at 1~nC is reduced by factor of $\sim2$ compared 
to the nominal-laser case when the homogenize laser is used; see simulated column in Table \ref{emittable} and Fig.~\ref{beamemit}.

\begin{figure}
\centering
 \includegraphics[width=0.89\linewidth]{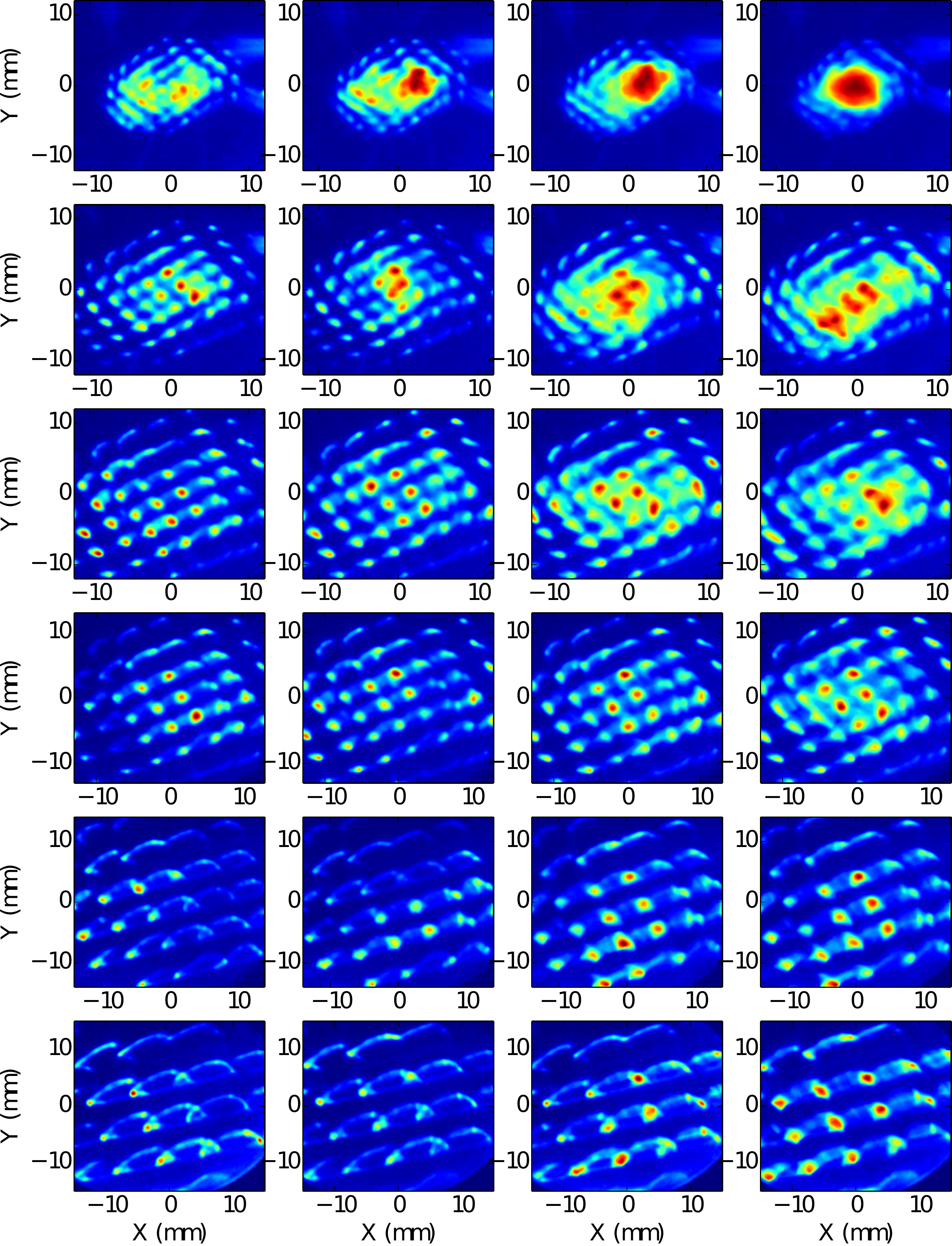}
  \caption{\label{YAG1}False color 7 MeV electron beam patterns for various matching solenoid current setting and charge.
From left to right: Q=60pC, 80pC, 100pC, 120pC. From top to bottom: M=215A, 230A, 241A, 255A, 270A, 290A. }
\end{figure}

The total maximum charge that still allows to observe the patterned beam at 
the position of the YAG1 was $\sim 1$~nC (see Fig. \ref{YAG1}). At higher charge the modulations tend to wash out due to increased space-charge effects. 

As a side note, we point out that the non cylindrical-symmetric  (square shaped) pattern
 is rotated due to the Larmor precession in the solenoids. It is therefore important to mount 
the MLA assembling on a rotatable optical stage for dynamics control of the final pattern angle. 
Such an approach would decouple the linac focusing (solenoid settings) from the requirement of 
preparing a pattern that leads to a strong horizontal modulation, e.g., for further injection 
in the transverse-to-longitudinal phase space exchanger. Additionally, this degree of freedom 
could allow to find rotation angles with higher-order bunching to reach higher modulation frequencies. 

\begin{figure}
 \includegraphics[width=0.98\linewidth]{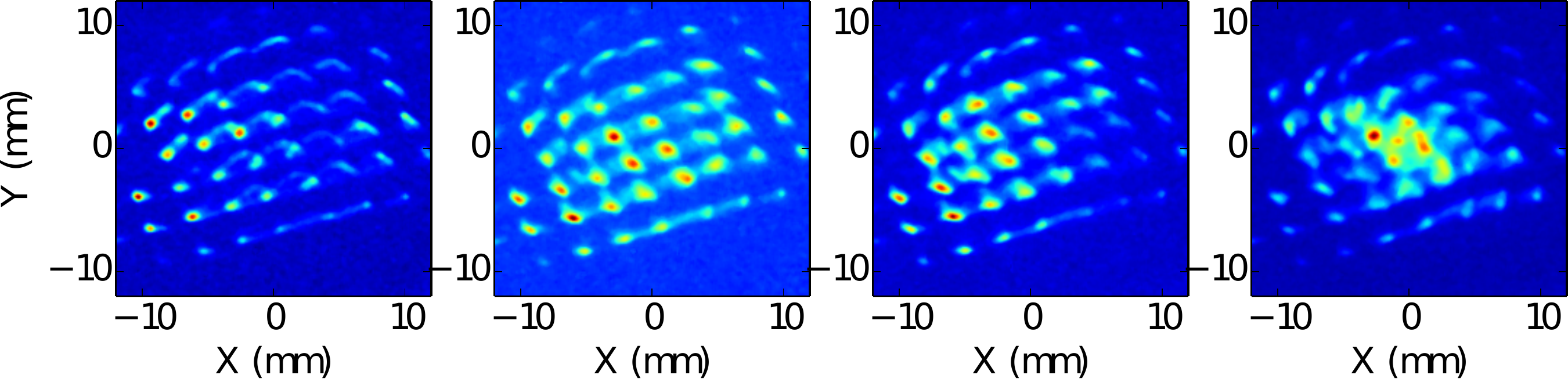}
  \caption{\label{YAG18}False color 18 MeV electron beam (gun + one linac) patterns for various charges.
From left to right: Q=60pC, 120pC, 140pC, 200pC. }
\end{figure}

\begin{figure}
 \includegraphics[width=0.98\linewidth]{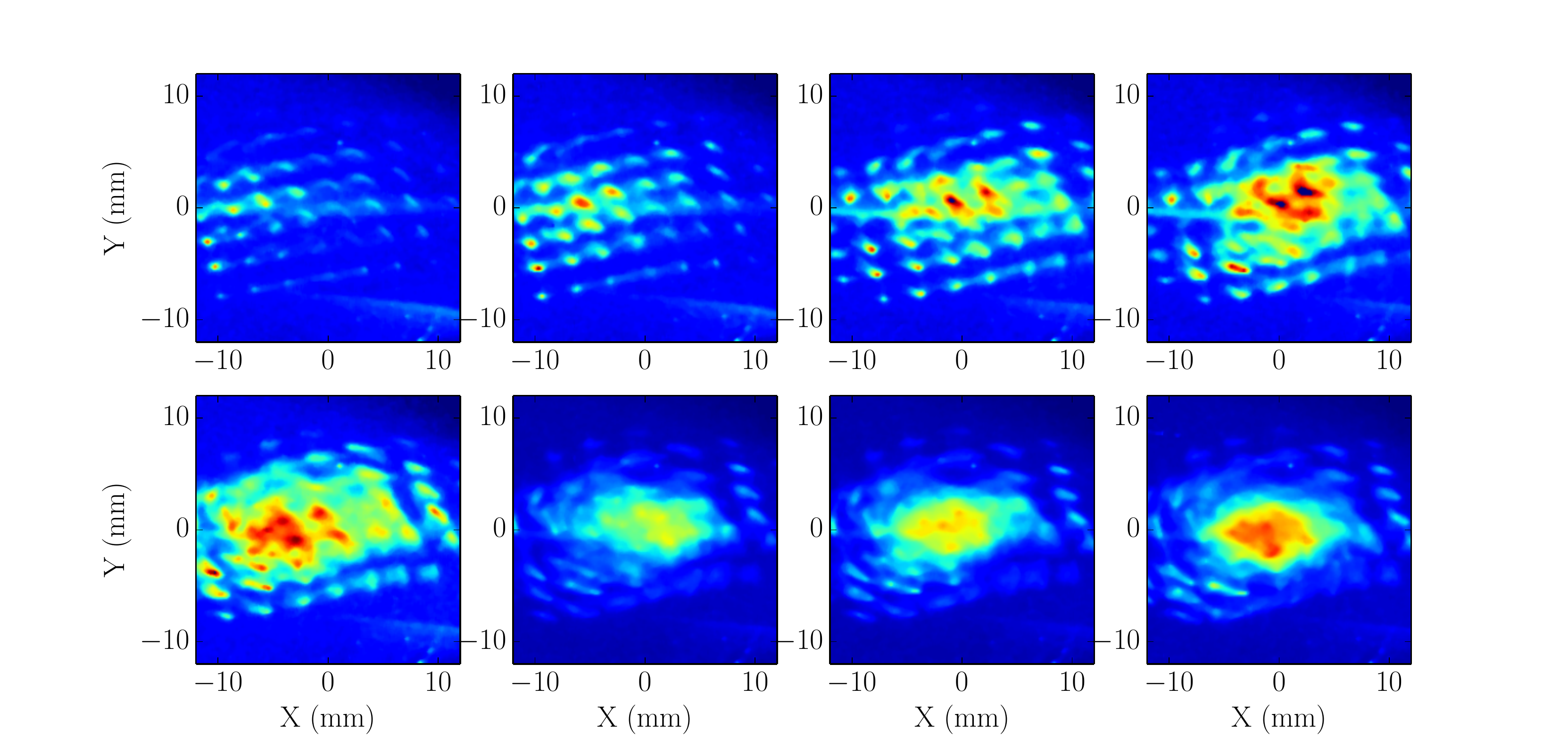}
  \caption{\label{YAG5}False color 50 MeV electron beam patterns for various charges. From left to right and top to bottom: Q=60pC, 100pC,
 200pC, 300pC, 400pC, 500pC, 600pC, 700pC. }
\end{figure}

\section{Low energy beam experiment}

In a first set of experiments we use a $8\times8$ laser-beamlet array to form a transversely-segmented 8-MeV electron beam. 
The corresponding electron-beam transverse images at the YAG1 ($z=3.07$~m from the photocathode surface) for different settings 
of the focusing-bucking
and matching solenoids appears in Fig.~\ref{YAG1}. Note, that due to the space charge effects close to the photocathode surface, 
the charge emission associated to each beamlet, and therefore the total charge of the patterned beam, is limited.  In our experiment 
the maximum total charge of the patterned beam was approximately $\sim 1.5$~nC or $\sim 30$ pC per beamlet.
 
\begin{figure}
\centering
\includegraphics[width=0.492\linewidth]{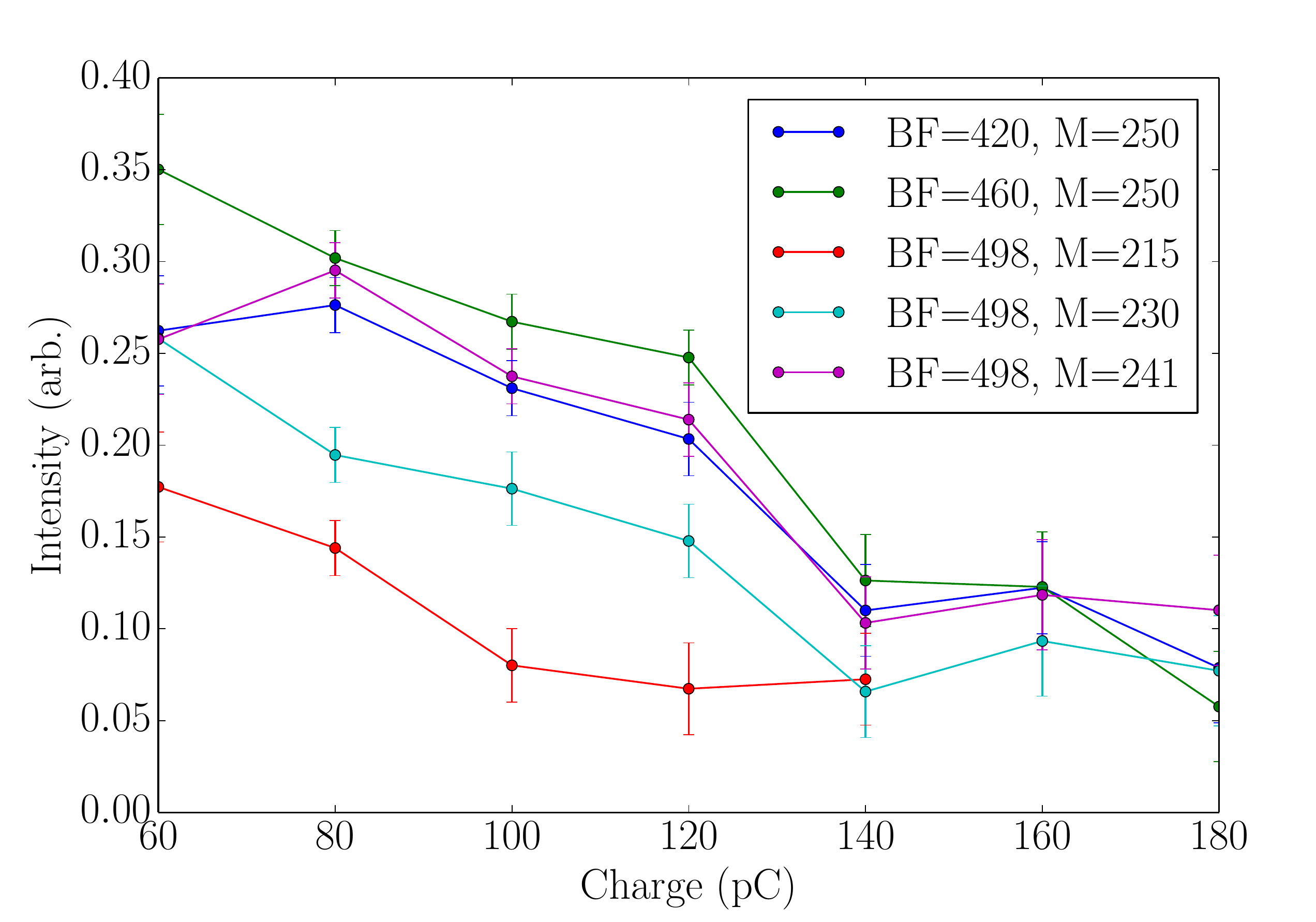}
\includegraphics[width=0.482\linewidth]{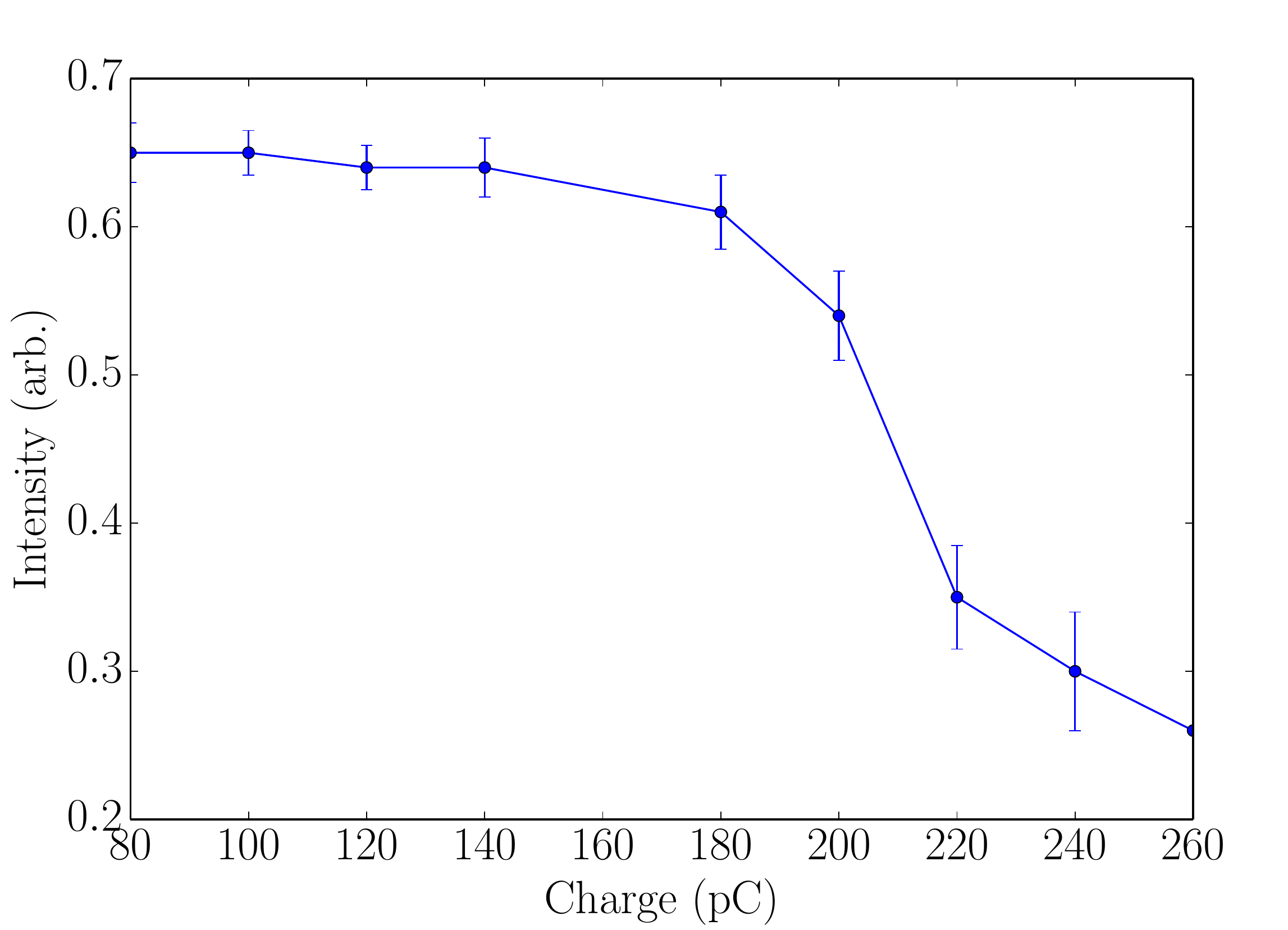}
\caption{\label{ebeammod}First harmonic amplitude vs. charge for different solenoid settings (left). First harmonic amplitude vs. charge in case of strong (M=290A) matching (right). }
\end{figure}

The electron beamlets formation is analyzed using the same Fourier analysis as the one used for  laser images. Figure ~\ref{ebeammod} represents the evolution of the transverse bunching factor versus total bunch charge for different solenoid settings. 
We observe that the transverse bunching decreases as the charge is increased. 
Additionally, we find that larger solenoid value tend to form ``tails'' due to energy spread in beamlets; see Fig.~\ref{YAG1}.
It should be noted, that this value is taken at YAG1 and does not provide information on possible modulation reappearance at a downstream position along the beamline.

\section{48 MeV beam experiment}
To preserve the beamlet pattern up to the emittance exchanger (EEX) entrance (see Fig.~\ref{beamline}, far left),
 there should be no strong focusing applied along the beamline as close encounter of the beamlet produces
 strong distortion as explored in Ref.~\cite{Rihaoui}. Consequently, the low-energy beamline  elements should be 
properly matched to allow the large waist. In the present experiment, we matched a 18 MeV electron beam onto YAG1 screen
as a first step; see Fig.~\ref{YAG18}.
 We used an additional solenoid (linac solenoid 1 on Fig.~\ref{beamline}) to image the beamlet formation on the YAG5 screen located 
just upstream of the EEX ($z=14.91$~m from the photocathode surface).

As one can see in Fig.~\ref{YAG5}, the beamlet separation is on the order of $\sim 2$~mm, which can be later optimized for
 generating THz transition radiation downstream of the EEX beamline~\cite{Ha:2016qrz}. As already noted, the coupling at 
YAG5 could be removed by mounting the MLA assembly on a rotatable mount. Likewise, the coupling could be taken advantage of to select an angle so that a smaller projected separation along the horizontal axis could be achieved. 
Such a configuration would provide a knob to continuously and promptly vary the beamlet separation (and THz-enhancement frequency) downstream of the EEX.

Additionally, an experiment to support the benefits of homogenizing the electron beam by 
setting up the MLA to produce an homogenized laser spot on the photocathodes was carried out. 
The beam was subsequently homogenized  (upper right picture in Fig.~\ref{noarray}), clipped
 with a circular aperture to yield the same rms sizes as the nominal laser spot, and its 
emittance was measured. The emittance measurement was made using a pepper-pot technique \cite{pepperpot,Zhang:1996af}.
A circular mask of 7.5~$\mu$m was inserted, followed by the screen at 81.5 cm. The resulting emittance measurements are combined in Table \ref{emittable}.

Our results are consistent with {\sc gpt}~\cite{GPT} simulations and indicate the homogenized 
laser reduces the transverse emittance by a factor $\sim 2$ compared to the nominal laser distribution. 
Prior to the experiment, no specific procedure was used to minimize the nominal emittances and the emittance values were rather large.

\begin{table}[]
\centering
\begin{tabular}{@{}|l|l|l|l|l|@{}}
\hline
\multicolumn{1}{|c|}{\multirow{2}{*}{Measurement}} & \multicolumn{2}{c|}{No MLA ($\mu m$)} & \multicolumn{2}{c|}{MLA ($\mu m$)} \\
\multicolumn{1}{|c|}{}                             &  $\epsilon_x$  & $\epsilon_y$  & $\epsilon_x$  &$\epsilon_y$ \\ 
\hline
GPT simulations (2D linac map)                              &     16.8        &    13.5    &    6.8    &    6.8     \\ 
\hline
Pepper pot (core)                                  &         8.2      &  6.5     &         6.3    &     4.9          \\ 
\hline
Pepper pot (full bunch)                                  &     13.1         &     11.6         &     9.2        &    8.7        \\ 
\hline
\end{tabular}
\caption{\label{emittable}Resulting emittance measurement for $Q=1$ nC.}
\label{my-label}
\end{table}

\section{Summary}
We demonstrated the possible use of MLA to control the transverse distribution of a photocathode laser pulse and associated photoemitted electron bunches. We especially 
confirm that this simple setup could significantly improve the transverse electron-beam emittance. Additionally,  we explored the generation of patterned beams consisting of multiple transversely-separated beamlets. The later distribution combined with the emittance exchanger beamline available at AWA could lead to the formation of bunch trains in the time domain~\cite{graves}. 

\section{Appendix A: Further MLA testing}

We attempted to measure the evolution of the first harmonic of the Fourier 
spectrum as a function of the Fourier lens position with respect to the MLA and the spacing
between the two MLAs. 
The production of modulated beam patterns appears to be possible for various
focal lengths as expected. We selected the Fourier lens focal length to be $f=250$~mm for convenience.

\begin{figure}[t]
\begin{center}
\centering
 \includegraphics[width=0.50\linewidth]{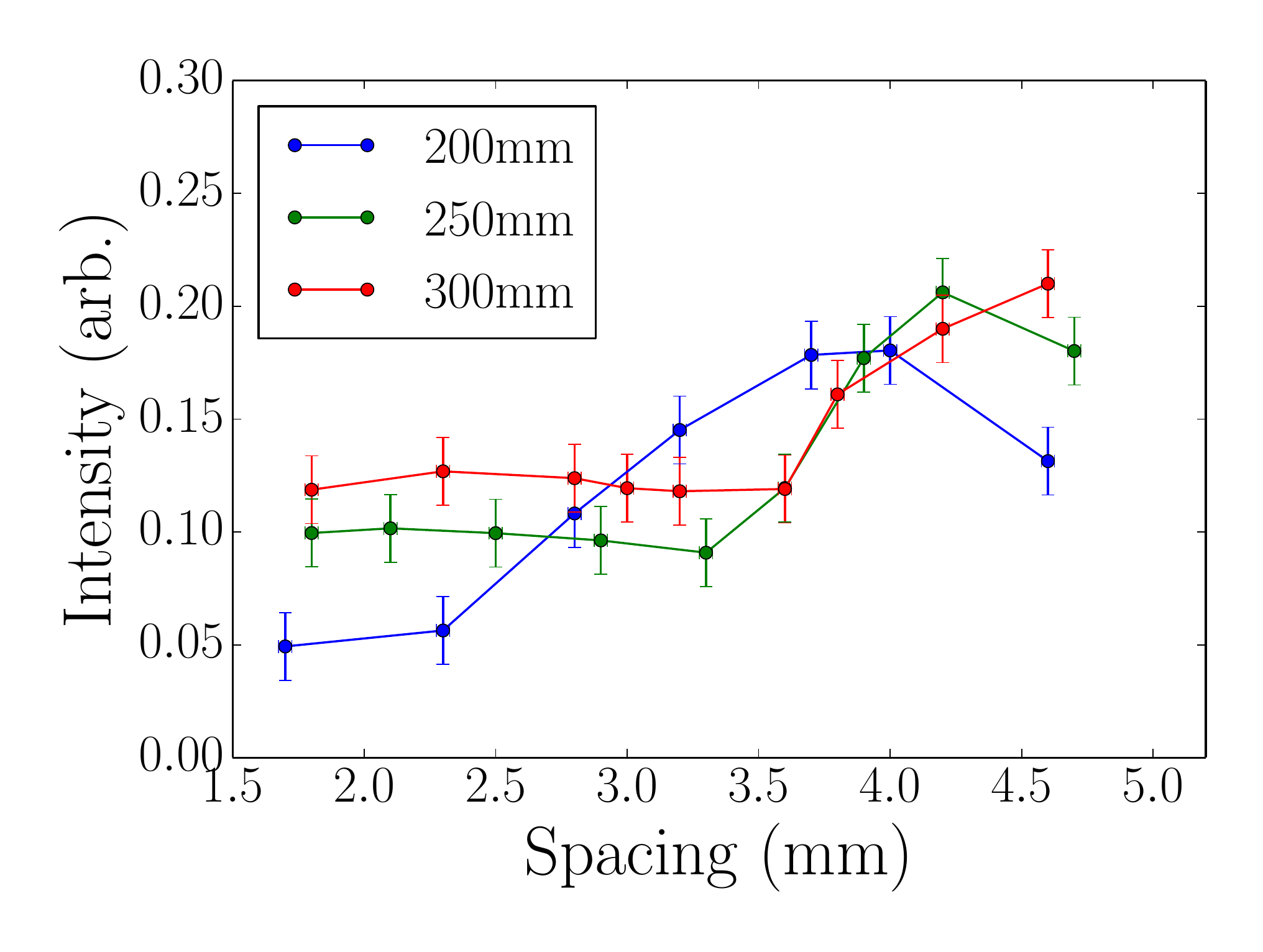}
\includegraphics[width=0.43\linewidth]{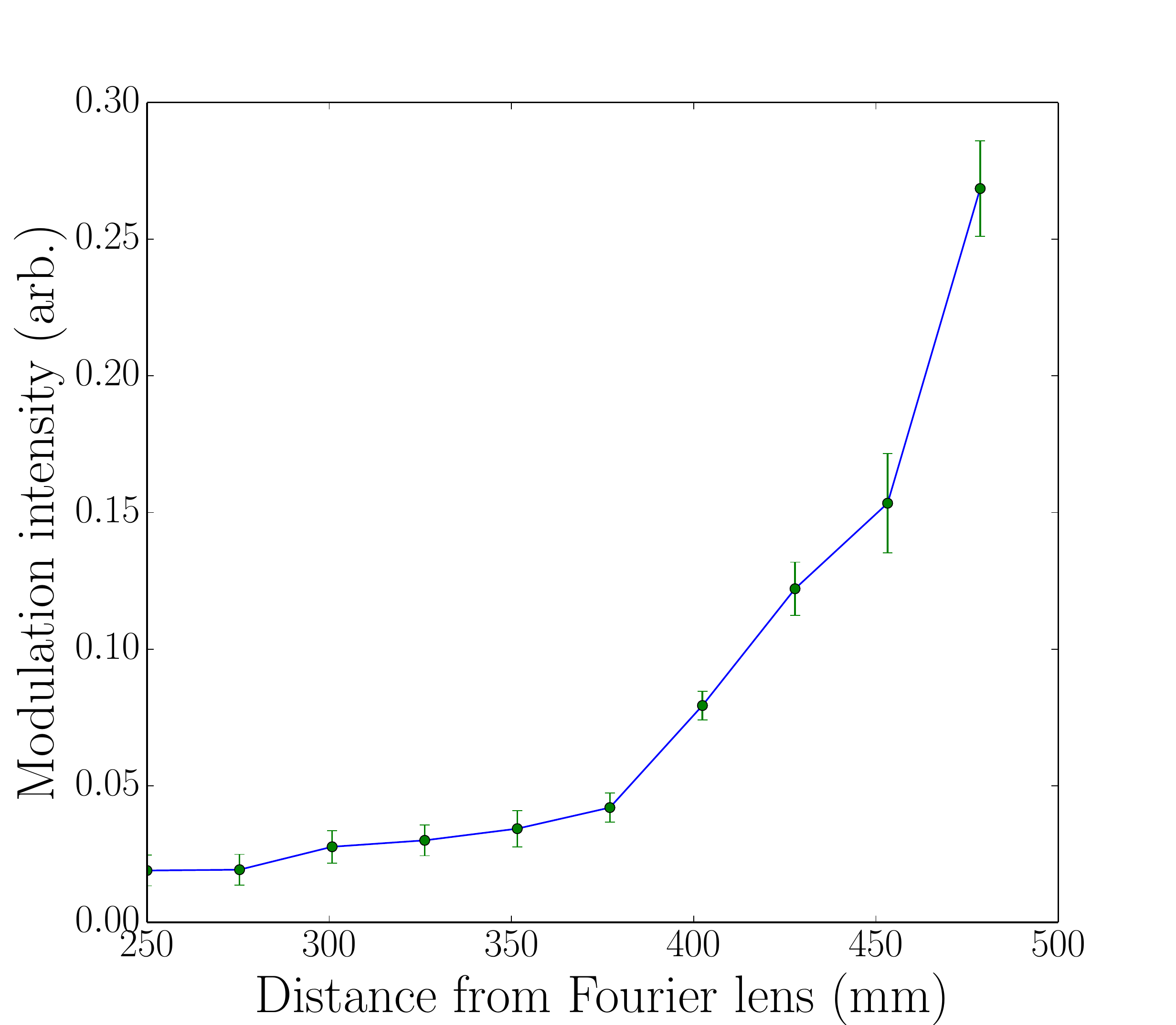} 
\caption{\label{intvssp}The intensity of the first harmonic vs. spacing between two arrays for different values of the focal length (left).
The intensity of the first harmonic as a function of the Fourier lens focal length (right).}
\end{center}
\end{figure}

We can conclude from these measurements that it is relatively easy to produce a
homogenized beam (as many of the studies conditions led to 
bunching factor at the 10-20\% level); see Fig.~\ref{intvssp}.  In Fig.~\ref{intvssp}, 
the bunching factor value was plotted against the spacing $d$ between two MLA plates. 
The modulation frequency is not reported in Fig.~\ref{intvssp}.

To study the behavior of the FFT spectrum as a function of Fourier lens position and to develop the quantification of the corresponding roughness of the image,
we varied the Fourier lens position between homogenization and modulation points. The resulting graph (see Fig. \ref{intvssp}) shows that for 
the $f=250$ mm there is a region of relatively low modulation (10-20\%) beyond the focal plane, which gives more flexibility
for use in real optical beamlines. 
Further optimization of the uniformity or bunching factor of the resulting image can be done by a motorized mount to control the spacing between the MLAs.

\section{Appendix B: Beamlet formation distortion analysis }

\begin{figure}[t]
\centering
\includegraphics[width=0.42\linewidth]{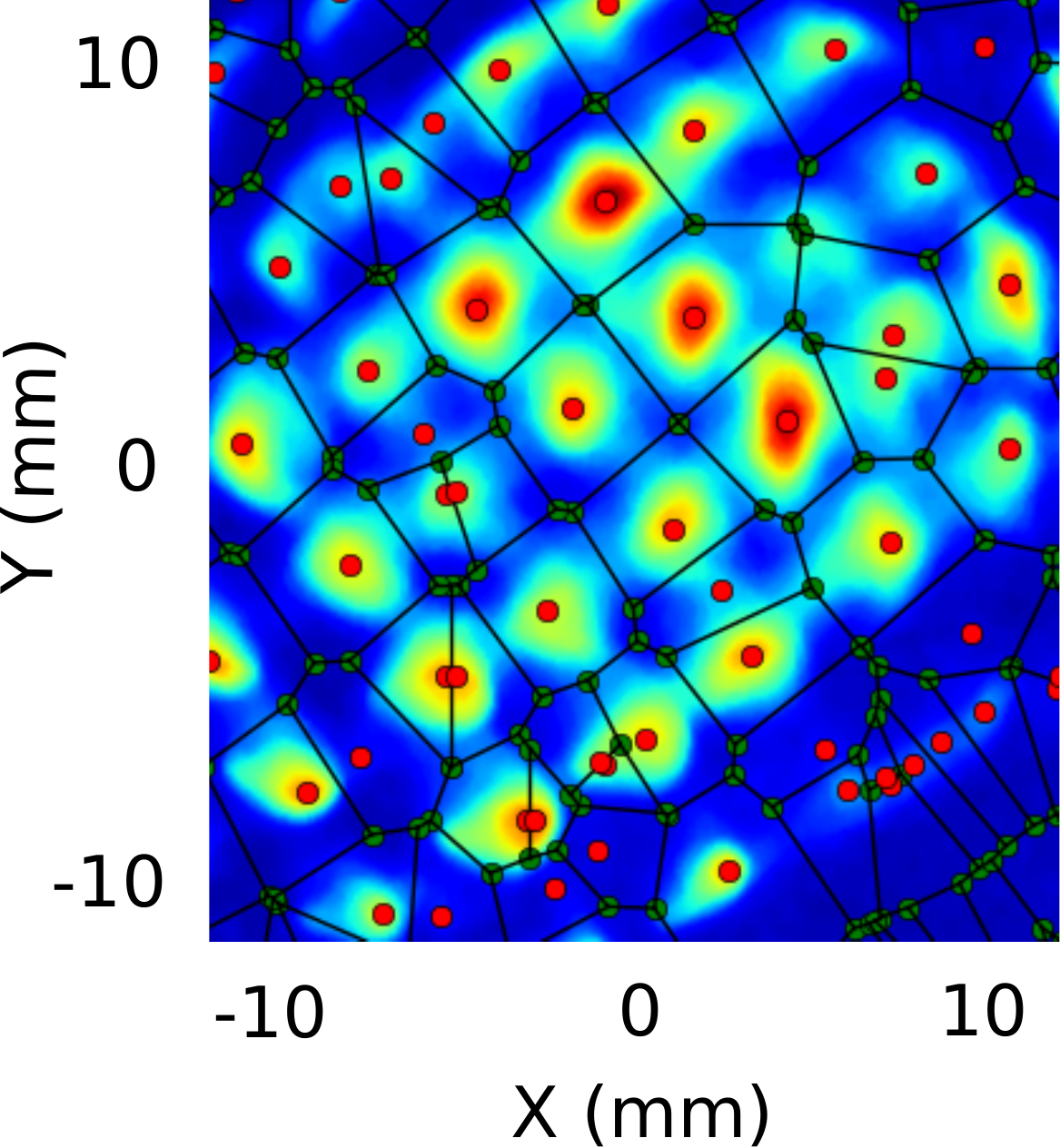}
 \includegraphics[width=0.53\linewidth]{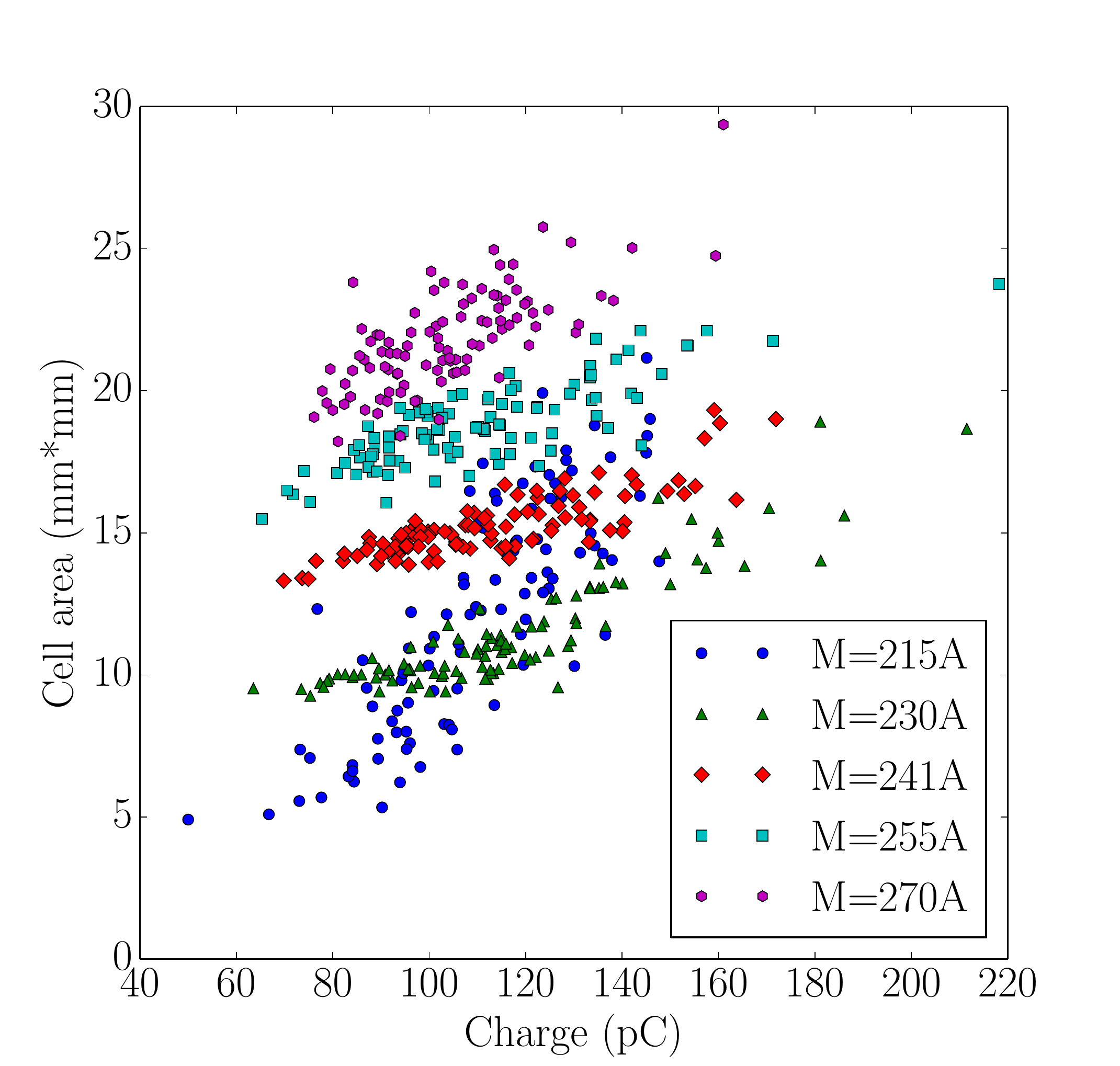}
  \caption{\label{celldata} Demonstration of Voronoi mesh generation for beamlet formation (left).
Average Voronoi cell area for different matching solenoid current as a function of charge (right). }
\end{figure}

To analyze the mean distortion of the beamlet pattern due to space charge effect and rf gun coupler kicks, we 
used Voronoi tessellation procedure available within SciPy package. 
This technique is inspired by galaxy clusters image analysis in Astrophysics \cite{Neyrinck}.

First, a pair of neighboring beamlets is connected with a line
segment and a perpendicular bisector is drawn onto that segment. When bisectors from different line segments intersect they form a
Voronoi cell vortex; see Fig. \ref{celldata} (left). In case of perfectly grid-lined beamlet formation each cell (excluding edge cells) will have exactly four
vortices and a rectangular shape with the side equal to the spacing between beamlets.  
When distortion to the formation is introduced, the cells become convex polygons, thus as a measure of distortion we
considered the average cell area.

At low matching solenoid currents the space charge force between beamlets is strong and the formation is almost smeared out at
charges above 80 pC. On Fig. \ref{celldata}, it corresponds to rapidly growing average cell area with charge.
With increase of the matching solenoid current the beamlets become more separated and at M=241A/255A the distortion of the initial formation
is reduced. This can be also visually confirmed by Fig. \ref{YAG1} (third/fourth row).
For the beamlet formation transport to 50 MeV experimental area at AWA we have chosen the medium matching solenoid current of M=240A.

With even higher matching current, the beamlets appear to have tails which are due to the chromatic effects in the solenoid.
This results in large spread of the cell area and the method becomes not applicable; see Fig.\ref{YAG1},\ref{simpat},\ref{celldata}.

\bibliography{biblio}

\begin{thebibliography}{10}

\bibitem{mlaabcd}
Peter Schreiber, Serge Kudaev, Peter Dannberg, and Uwe~D. Zeitner.
\newblock Homogeneous led-illumination using microlens arrays, 2005.

\bibitem{dickey2000laser}
F.M. Dickey and S.C. Holswade.
\newblock {\em Laser Beam Shaping: Theory and Techniques}.
\newblock Optical Science and Engineering. Taylor \& Francis, 2000.

\bibitem{deOliveira}
Otávio~Gomes de~Oliveira and Davies~William de~Lima~Monteiro.
\newblock Optimization of the hartmann–shack microlens array.
\newblock {\em Optics and Lasers in Engineering}, 49(4):521 -- 525, 2011.

\bibitem{asta}
Elvin Harms, Jerry Leibfritz, Sergei Nagaitsev, Philippe Piot, Jinhao Ruan,
  et~al.
\newblock {The Advanced Superconducting Test Accelerator at Fermilab}.
\newblock {\em ICFA Beam Dyn.Newslett.}, 64:133--156, 2014.

\bibitem{manoel}
et.~al. M.E.~Conde.
\newblock {\em Proceedings of IPAC'10}, 2010.
\newblock Kyoto, Japan, 4425.

\bibitem{sussinfo}
SUSS MicroOptics.
\newblock Smo techinfo sheet 10 - beam homogenizing.
\newblock 2008.

\bibitem{FZhou}
F.~Zhou, I.~Ben-Zvi, M.~Babzien, X.~Y. Chang, A.~Doyuran, R.~Malone, X.~J.
  Wang, and V.~Yakimenko.
\newblock Experimental characterization of emittance growth induced by the
  nonuniform transverse laser distribution in a photoinjector.
\newblock {\em Phys. Rev. ST Accel. Beams}, 5:094203, Sep 2002.

\bibitem{Rihaoui}
M.~Rihaoui, P.~Piot, J.~G. Power, Z.~Yusof, and W.~Gai.
\newblock Observation and simulation of space-charge effects in a
  radio-frequency photoinjector using a transverse multibeamlet distribution.
\newblock {\em Phys. Rev. ST Accel. Beams}, 12:124201, Dec 2009.

\bibitem{Astramanual}
K.~Floettmann.
\newblock Astra reference manual.

\bibitem{GPT}
et.~al. S.B. van~der Geer.
\newblock 3d space-charge model for gpt simulations of high brightness electron
  bunches.
\newblock {\em Institute of Physics Conference Series}, 175:101, 2005.

\bibitem{Barnes:1986}
J.~Barnes and P.~Hut.
\newblock A hierarchical o(n log n) force-calculation algorithm.
\newblock {\em Nature}, 324:446 -- 449, Dec 1986.

\bibitem{Halavanau:NIMA}
A.~Halavanau and P.~Piot.
\newblock Simulation of a cascaded longitudinal space charge amplifier for
  coherent radiation generation.
\newblock {\em Nuclear Instruments and Methods in Physics Research Section A:
  Accelerators, Spectrometers, Detectors and Associated Equipment}, 819:144 --
  153, 2016.

\bibitem{Ha:2016qrz}
Gwanghui Ha et~al.
\newblock {Demonstration of Current Profile Shaping using Double Dog-Leg
  Emittance Exchange Beam Line at Argonne Wakefield Accelerator}.
\newblock page TUOBB01, 2016.

\bibitem{pepperpot}
J.~G. Power, M.~E. Conde, W.~Gai, F.~Gao, R.~Konecny, W.~Liu, Z.~Yusof,
  P.~Piot, and M.~Rihaoui.
\newblock Pepper-pot based emittance measurements of the awa photoinjector.
\newblock pages 4393--4395, June 2007.

\bibitem{Zhang:1996af}
Min Zhang.
\newblock {Emittance formula for slits and pepper pot measurement}.
\newblock 1996.

\bibitem{graves}
W.~S. Graves, F.~X. K\"artner, D.~E. Moncton, and P.~Piot.
\newblock Graves \textit{et al.} reply:.
\newblock {\em Phys. Rev. Lett.}, 111:019402, Jul 2013.

\bibitem{Neyrinck}
Mark~C. Neyrinck.
\newblock zobov: a parameter-free void-finding algorithm.
\newblock {\em Monthly Notices of the Royal Astronomical Society},
  386(4):2101--2109, 2008.

\end{thebibliography}
\bibliographystyle{unsrt}

\end{document}